 \documentclass[final,3p,times]{elsarticle}

\usepackage{epsfig}

\usepackage{amssymb}

\journal{International Journal of non-linear mechanics}

\begin{document}

\begin{frontmatter}



\title{Theory of ice-skating}



\author {Martine Le Berre$^{1}$ and Yves Pomeau$^{2}$ }
\address{$^1$Ismo (CNRS UMR 8214), Université de Paris-Saclay,
91405 Orsay, France}
\address{$^{2}$Ladhyx (CNRS UMR 7646), Ecole Polytechnique, 91128 Palaiseau, France}

\begin{abstract}
\label{Abstract}
Almost frictionless skating on ice relies on a thin layer of melted water insulating mechanically the blade of the skate from ice. 
Using the basic equations of fluid mechanics and Stefan law, we derive a set of two coupled equations for the thickness of the film and the length of contact, a length scale which  cannot be taken as its value at rest. The analytical study of these equations allows to define a small a-dimensional parameter depending on the longitudinal coordinate which can be neglected  everywhere except  close to the contact points at the front and the end of the blade, where  a boundary layer solution is given. This solution provides without any calculation the order of magnitude of the film thickness,  and its  dependence with respect to external parameters like the velocity and mass of the skater and the radius of profile and bite angle of the blade, in  good agreement with the numerical study. Moreover this solution also shows that a lubricating water layer of macroscopic thickness always exists for standard values of ice skating data, contrary to what happens in the case of cavitation of droplets due to thermal heating (Leidenfrost effect). .
\end{abstract}


\end{frontmatter}

\date{\today}


\section*{Foreword written by Yves Pomeau}
\label{foreword}
The present paper reports original work which seems to me
 a very appropriate topic for this volume honoring Martine Ben Amar. Its matter is related to my first (very successful) collaboration with her. Our first joint paper \cite{joint}
studied the growth of needle crystals in an undercooled melt. At the time it was well understood that the classical Ivantsov solution for the needle crystal had to be supplemented by physical effects outside of the ones in Ivantsov theory to yield a solution which would be both unique and pertinent for the observations. It had long been suspected that surface tension, through the Gibbs-Thomson curvature-dependence of the equilibrium temperature, had to be taken into account. But, at the time, it was more or less an article of faith. With Martine, we did show  that things did not work as believed by many in the field: no solution exists with isotropic solid-melt surface tension. Crystal anisotropy had to be taken into account  \cite{joint}. This advance was followed by many others thanks to Martine and her collaborators so that this tricky problem can now be considered as solved. 

The connection with the present work is that, as in the 1986 Europhys. Lett paper, we have to introduce 
Stefan condition for melting to understand quantitatively how a supply of heat can explain the dynamics of the solid/liquid interface. Ice skating relies on the same dynamics, and as the literature shows too well, it is a quite non trivial matter to write the appropriate boundary conditions. It is quite amazing to come back to this "classical", but somewhat forgotten, physics after so many years. 

I wish you Martine, many more years of active and fruitful research.

\section{Introduction}
\label{intro}
Skating is possible because of the very small friction felt by a thin blade sliding on ice, in sharp contrast with much larger solid-on-solid friction observed at temperatures well below the melting temperature. This has long been explained \cite{reynolds} by the existence of a thin lubricating layer of liquid water. Standard wisdom was that ice melts because of the local increase of pressure by the weight of the skater, a pressure leading to local melting because of the pressure dependence of equilibrium melting temperature. But this melting is an equilibrium phenomenon, then it is unclear how it can describe the continuous formation of liquid water from ice,  in particular because this requires a supply of heat. According to another explanation \cite{A} the main phenomenon is melting by the heat generated by friction in-between ice and the sliding skate. This heat yields the energy needed to balance the latent heat, as given by Stefan  condition on the ice surface. A condition of mechanical equilibrium expresses also how the weight of the skater is supported by the pressure generated inside the film, which avoids direct ice-skate contact.

We reconsider below this question in the light of a macroscopic approach, assuming that the liquid layer in-between ice and skate has macroscopic thickness. Even though this layer is often referred to as a lubricating layer, it acts differently of a regular lubricant: lubricants like oil are chemically different of the solids facing each other. Therefore a layer of molecular thickness  can change the conditions of sliding. On the contrary, in the present case, the lubricating layer is made of the same molecules of water as solid ice, so that a layer of molecular thickness, which is about one angstrom for water, will be mixed with the molecules present at the surface of ice, and cannot change the ice-skate friction.

Our analysis is inspired by previous work on the Leidenfrost effect \cite{leiden} where a thin layer of vapor was also analyzed and where the balance of vertical forces  (weight of the airborne Leidenfrost drop and pressure forces) plays a central role.  A difference between the skating problem and the Leidenfrost effect is that in Leidenfrost the evaporation from the droplet (instead of the melting in skating) is due to the heating by the hot bottom plate although in skating the heat comes from viscous friction in the bulk of the lubricating layer. Here it is the longitudinal viscous friction (along the direction of motion) which is responsible for heating the film of water, although this longitudinal flow does not generate any vertical stress in the layer. The weight of the skater is actually supported by the transverse Stefan flow which undergoes a drop of pressure bigger than in the longitudinal direction in the skating case (this is not the case treated  in \cite{A}, which considers a 2-D situation).

We answer a very natural question: when a sharp edge slides on ice, how deep is the furrow due to melting arising itself from the heat generated by viscous friction? This introduces a length scale (the depth of the furrow) which turns out to be much bigger than the thickness of the liquid layer due to melting, a layer in-between ice and skate. To estimate this contribution of melting one needs to find out how much water is made during the time a skate passes over a point on the ice surface, and so how much ice has melt to make the furrow. We show that this  "lubricating" layer has macroscopic thickness in concrete situations, but that it is far thinner than the depth of the furrow, an important remark. Furthermore, as written above, the depth of the furrow is far smaller than its length which is far smaller than than the radius of profile of the blade. Hence one has to solve a problem with four different length scales, each one with a different order of magnitude than the the three others, see equation (\ref{eq:inegalfa}) referring to the lengths reported in Fig.\ref{fig:sec3}.

Here we present a theory using  in a consistent way Stefan description of the melt/solid interface. 
This is not a trivial endeavor, because one has to write the equations for fluid mechanics in the frame of reference where the boundaries of this film are fixed, namely in the frame of reference of the skater. Therefore when writing Stefan condition on the surface of ice, we must consider the interface as a moving surface with respect to ice \textit{and}  the ice as moving  with respect to the skater. It follows that the ice melting rate cannot be identified to the growth rate of the film thickness, as done in the recent model derived by  Lozowski et al.\cite{LScana}-\cite{LScana2}. We show below that the missing term in their model is actually of prime importance.

By combining  (i) Stephan condition with (ii) the balance between the weight of the skater and the viscous pressure force in the liquid layer, and (iii) the relation between the volume of the trough and the melting process during the passage of the skater, we derive two coupled integro-differential equations relating the  film thickness to the length of contact. All physical lengths  of the problem are then deduced from the data,  in particular  the length of contact between ice and skate, a parameter which cannot be set to its value at rest as done in \cite{LScana}-\cite{LScana2}.
 The study of these coupled equations allows to define a small a-dimensional quantity depending on the longitudinal coordinate (in the direction of the motion) which can be neglected  everywhere except  close to the contact points at the front and the end of the blade. Help to this property we show that the length of contact can be deduced from the integral equation with a very good approximation. Then it remains to solve a single differential equation for the film thickness.  From the  analytical study of the two boundary layers at the front and at the end of the blade, we show that one can deduce an expression of the order of magnitude of the film thickness in these two regions in terms of the data. This provides without any calculation the relation between the unknown variables and the external parameters like the velocity and mass of the skater and the radius of profile and bite angle of the blade, in  good agreement with the numerical study.
 
In section \ref{sec:hockey} we derive the model equations for standard ice-skating conditions, namely for V-shaped blade transverse profile, and study in details the solution analytically and numerically with applications to hockey and inclined speed- skating blades. Because of the existence of a small parameter we define a critical mass below which no skating is possible. This critical mass is so small that one may conclude that a macroscopical film of water is  always formed in the  skating case, a conclusion not valid for cavitation of water droplet in the Leidenfrost effect \cite {leiden}. The case of a vertical speed-skating or rectangular transverse blade profile is treated in section \ref{sec:rectangular} which gives results  very similar to the ones for V-shaped blades, although with mote complex analytical  expressions. Last section is devoted to conclusions. This version rectifies an error (on the value of the water density\footnote{we are greatful to Christophe Clanet for having pointed out this  mistake.})  which was present  in the previous version  already published in the International Journal of non-linear mechanics.

\section{ Equations for V-shaped blades}
\label {sec:hockey}
\begin{figure}
{
(a) \includegraphics[height=1.1 in]{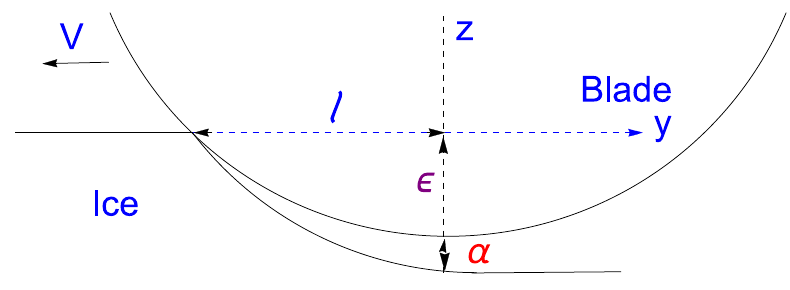}
(b) \includegraphics[height=1.15in]{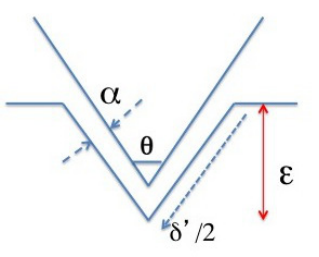}
}
\caption{ Schema of the blade digging into the ice surface. In (a) we set $y=0$ at the front contact point and $y=\ell$ at the back contact point where the letters $\epsilon$ and $\alpha$ are placed.  (b) Schema of the cross-section at a given value of $y$. The geometry is deformed in order to see the various lengths.
}
\label{fig:sec3}
\end{figure}

To get close to  ice-skating conditions, we consider first a skate having a smooth surface ending toward the ice with a sharp edge, its longer dimension being in the direction $y$ of the imposed speed, see fig.\ref{fig:sec3}.  Such V-shaped blades are the ones sharpened for hockey skaters and also figure skatings. The bottom of each blade present  actually  a hollow separating two V-shaped edges but the skater spends most of his time on a single edge, therefore we focus here on the case of  V-shape transverse profile on contact with the  ice surface via a thin film of melted water. In this geometry the water layer has only a finite extend along the horizontal coordinate $x$ perpendicular to the direction of motion, a width much smaller than the length of the skate.  Lastly the vertical direction is associated to coordinate $z$. The curved profile of the blade and the cross-section  are schematized in figs.\ref{fig:sec3} where the length scales are not respected. The length of contact $\ell$ is defined as usual, as the longitudinal dimension of the wetted part of the blade which is situated between the first point of contact (where the thickness of the layer is zero) and the point of maximum penetration of the blade into the ice. Beyond this point the contact is assumed to stop because the bottom of the curved blade becomes above the surface of the through dug by the passage of the skater. 
This geometry also describes the case of speed skating (having rectangular cross-section) with inclined blade making an angle $\pi/4$  with the vertical axis, considered in subsection \ref{sec:numerics}.    


We derive below the equations for the flow created by the friction forces along $y$ which is partially evacuated laterally by a secondary flow flowing in the $x$ direction (perpendicular to the plane of Fig.(a))  generating a pressure supporting the weight of the skater. Note that there is also some flow ejecting fluid behind the blade which is generated by the pressure gradient along $y$. This gradient is negligible with respect to the lateral one because the length of contact along $y$ is much larger than the transverse dimension of the trough created by the passage of the blade.  We neglect the imperfections of sharpened wedges, and the fact that sharpening of the blade leaves a slightly rounded cylindrical wedge with a very small radius.  The radius of  curvature  of this wedge at the bottom of the blade (in the plane perpendicular to $y$)  is generally of the order of one micrometer, that is comparable to the film thickness found below.  But the  domain  where this hypothesis may be invalid (in  the  very bottom of the blade) is much smaller than the total water film domain (which includes the sides ploughing into the ice), therefore our  partially incorrect approximation  is  not really questionable for our purpose which is to derive the geometry of the furrow whose main part has length scales much larger than the radius of curvature or the imperfections of sharpened wedges.

\subsection{Derivation of the model  with Stefan condition}
\label {sec:stefan sol}

Below we write Stefan equation which links the velocity of the flow (or melting rate, normal to the ice surface) and the film thickness via the viscosity, the  surface in contact and the  latent heat of the ice. This equation set the balance between the heat generated by viscous friction in the liquid layer of melt water separating solid ice from the skate and the latent heat melting ice. This viscous friction is due to the fact that the layer has a small thickness $ \alpha$ and that across this thickness the $y$-component of the velocity goes from $V$ on the surface of the skate to zero on solid ice. This yields a viscous stress and so a source of heat in the liquid layer. For the moment we assume that all this heat is used to melt ice. 

Consider the  slice  $y,y+dy$. The film of water in this slice has a volume $\delta'(y) \alpha(y) dy$, where $\alpha(y)$ is the thickness of the film and $\delta'(y)$ is the length  of the wetted part of the blade, represented in Fig.\ref{fig:sec3}-(b) as the length of the two sides of a triangle. The power dissipated by friction in the film of water of volume $\delta'(y) \alpha(y) dy$ is equal to
$$dP=\eta \frac{V^2}{\alpha}\delta'(y)dy$$ where $\eta$ is the shear viscosity of water. 
In the cross section we define local coordinates $(s,n)$, not represented in the figure, with $-\frac{\delta'}{2} \le s \le +\frac{\delta'}{2}$ along the ice surface, and $ 0 \le n \le \alpha$ perpendicular to the surface, $n=\alpha$ on the skate. We assume that the thickness $\alpha(y)$ of the film is independent of $s$ in a given cross-section (an assumption consistent with the rest of the calculation) and that 
\begin{equation}
\alpha(y) \ll \epsilon \ll \ell \ll r,
\label{eq:inegalfa}
\end{equation}
where $r$ is the radius of profile of the slightly curved blade which is of several meters or more (the curvature is intentionally exaggerated in the schema (b)).
This condition implies
$\alpha << \delta'$, because the bite angle $\theta$ is typically around $\frac{\pi}{2}$. This power $dP$ is responsible for the melting of a volume $ d\mathbb{V} =w_0(y)\delta'(y)dy$ , where $w_0(y)$ is the melt rate. We assume that $w_0(y)$  is independent of the local coordinate $s$.
The power $dP$  dissipated by the shear stress is equal to the power required to melt the volume $ d\mathbb{V}$, then one has $dP= L \rho_{s} d\mathbb{V}$  where $ \rho_{s}$ is the density of the ice (at the surface). It gives Stefan equation valid when the heat losses are neglected (for perfectly insulating blades)
 \begin{equation}
\alpha(y) \vert w_0(y) \vert =\eta \frac{V^{2}}{ L \rho_s}.
\label{eq:wo1a}
\end{equation}

In the following we shall neglect the difference of density between ice and water to simplify the analysis and put $\rho_{s}=\rho_{l}$, a minor approximation in regards with others (like the assumption of  constant film thickness in the cross-section which will be assumed below), our aim being to get in fine an order of magnitude of the geometry of the trough and of the film, and ultimately to define the conditions for skating with small friction.
Note that the melt rate $w_{0}$  cannot be identified to $d\alpha /dt$, see the discussion below equation (\ref{eq:wosk}).

 \subsubsection{Heat transfer in the film}
 
 The equation (\ref{eq:wo1a}) is a possible writing of Stefan condition. It expresses the balance between the power input coming from the viscous stress and the latent heat needed to melt ice. This is an approximation only, because it short cuts the heat transfer problem in the liquid layer. If the matter of the skate had zero heat conduction, this formula were exact, because the only place the heat due to viscous friction could go is to ice in order to melt it. Consider the opposite limit where the matter of the skate is a perfect heat conductor. In this case one has also to assume something about the temperature of the skate. To simplify the matter let us take it as equal to the melting temperature. In this case, the temperature in the liquid layer will be symmetrical with respect to the mid plane of the layer, so that the heat flux toward the surface of the skate will be the same as the one toward the ice. In this case only half of the heat generated by viscous friction will be used to melt ice. 
 
 Moreover there is another possible effect, due to ice losses, because ice has to be heated to the melting temperature before to melt. As well known since Nansen \cite{Nansen}, this has a strong influence on sliding if ice is at very low temperature: then the viscous friction has first to heat ice and then to melt it. Because of the quite large latent heat of ice, this is a negligible effect in the condition of ice-skating in arenas, where the temperature of ice is just a little below melting. Therefore in this case, one can assume that all the heating is used to melt ice, except of course that it is also used to heat the skate itself. In the case of a skate conducting heat much better than water, as argued above only half of the heat is used to melt ice. 

In the following we consider 
a blade made of perfect heat conductor, that amounts to divide by $2$ the quantity of heat  generated by the work of viscous forces and used for melting.  With $ \rho_s=\rho_{l}=\rho$, Stefan  relation (\ref{eq:wo1a}) becomes
 \begin{equation}
\alpha(y) \vert w_0(y) \vert =\eta\frac{V^2}{ 2L\rho }.
\label{eq:wo1}
\end{equation}

 \subsubsection{Poiseuille flow description}
Neglecting the inertia of the flow (assumed to be at low Reynolds number), we can use the analytical solution of a Poiseuille flow. At given $y$, the gradient of the pressure $p$ in the film along the coordinate $s$ obeys the equation
 \begin{equation}
\frac{\partial p}{\partial s}= \eta \frac{\partial^2 u}{\partial n^2}, 
 \label{eq:dp1}
\end{equation}
where $u$ is the component of velocity in the film along $s$. Equation (\ref{eq:dp1}) has to be solved with the boundary conditions $\left(\frac{\partial u}{\partial n}\right)_{n=\alpha/2} =0$, $u(0)=u(\alpha)= u_{0} = -V\cos(\theta/2) \frac{\partial \epsilon_{sk}}{\partial y}$  which is due to the geometry of the skate  sunk into the furrow, $\epsilon_{sk}$  being the penetration depth of the blade at given $y$. The solution is

 \begin{equation}
u(s,n)= \frac{1}{2\eta}\frac{\partial p}{\partial s}(n^2-n\alpha) +u_{0}.
 \label{eq:u1}
\end{equation}

The components $u$ and $w$ of the velocity in the cross-section are linked by the condition of incompressibility ($w$ is the component in the direction $n$, normal to the surface of the blade). Assuming that the depth of the trough (of order $\delta'$) and the thickness $\alpha$ of the film are much smaller than the length of contact $\ell$, we can neglect the gradient of the pressure in the direction $y$ of the skater motion as compared to the two other components of this gradient. The conservation of mass condition reduces to

 \begin{equation}
\frac{\partial u}{\partial s}+ \frac{\partial w}{\partial n} =0.
 \label{eq:u2}
\end{equation}
Inserting the solution (\ref{eq:u1}) in (\ref{eq:u2}) and integrating along $n$ in the film, we obtain

 \begin{equation}
\frac{\partial^2 p}{\partial s^2}=  \frac{12 \eta\Delta w(y)}{\alpha^3},
 \label{eq:d2p}
\end{equation}

 where 
 
 \begin{equation}
\Delta w(y)=   w(y,n=\alpha) -w(y,n=0)
\label{eq:wosk}
\end{equation}

 results from the integral over $n$ of the Stefan flow. The expression of $\Delta w(y)$ deserves some comments and requires to explain  the meaning of the different velocities under consideration.
 
 Generally Stefan condition expresses a constraint on the fluid velocity normal to an interface where there is a phase transformation. There is a subtle point as this condition depends if it is written in the frame of reference of the interface or in the frame of the solid that is melting. To simplify the discussion let us consider the case, approximately true for ice melting, where there is no change of density from solid to liquid, as assumed here. Let us put oneself in the frame of reference where neither liquid water nor solid ice move, but where ice melts so that the surface of separation moves at speed $w$ with respect to the frame of both ice and water (which is possible if one neglects their density difference). In this frame the fluid velocity normal to the surface of ice will be zero, because the phase change does not transport any matter. The consequence of that is that, in the frame of reference of the ice surface, which is moving at speed $-w$ with respect to the frame of reference just considered, the fluid velocity near the surface must be also $w$ because of the Galilean transform necessary to put the fluid in the moving frame of reference of the ice surface. Hence in this frame one must impose a Stefan condition to the normal velocity $w$ of the fluid on the ice surface, or  with our notation $ w(y,0)= w_{0}(y)$.
 
On the contrary, on the surface of the skate, there is no phase transformation and the boundary condition in the frame of reference of the skate is the regular condition (for a viscous fluid) that all components of the fluid speed vanish on the surface of the skate in the frame of reference where it is at rest. Finally, in the skate frame, equation (\ref{eq:wosk}) writes

 \begin{equation}
\vert \Delta w(y) \vert= w_{0}(y),
\label{eq:deltaw}
\end{equation}
 
a quantity which was defined as the rate of melting and  cannot be identified to $\frac{d\alpha}{dt}$, see relation (\ref{eq:wo2}). 
 
Assuming the right-hand side of (\ref{eq:d2p}) independent of $s$, integrating  twice over $s$  and taking into account the boundary conditions $(\frac{\partial p}{\partial s})_0=0$ and $p(\delta'/2)=0$ we obtain 
the drop of pressure along the blade in the cross section

 \begin{equation}
p(s)= \frac{3}{2}\eta (\Delta w/\alpha^3)((\delta'_{sk})^{2} - 4s^2),
\label{eq:pr}
\end{equation}

where  $\delta'_{sk}$ is the  length of the wetted part of the blade in the cross-section, equal to
 $\frac {2 \epsilon_{sk} }{\cos(\theta/2)} $,  
$\epsilon_{sk}$ being given as a function of $y$  in equation (\ref{eq:epsk}).
 
 Along the vertical direction  $z$ the weight of the skater must be the opposite of the integral over $s$ of the vertical component  of the pressure. This yields  the condition
 
 \begin{equation}
\int_{0}^{\ell}{ \eta \sin(\theta/2){ \Delta w(y)}  \left(\frac{\delta'_{sk}(y)}{\alpha(y)}\right)^3}dy = M g ,
\label{eq:poids}
\end{equation}

where $\ell$ is the length of contact, $M$ the mass of the skater sliding on a single wedge of the blade (and assuming that its weight is supported by a single skate) and $g$ the acceleration of gravity. 

Now we have to describe the formation of the trough, that provides a relation connecting $w_{0}$ and $ V \frac{ \partial {\epsilon}}{ \partial {y}}$ . The  total area of the melted water in the V-shaped cross-section at $y$  is  equal to $\frac{1}{2} \epsilon(y)\delta(y)$. This area results from the passage of the skater and the ensuing melting during the time $\delta t= y/V$. Therefore 
\begin{equation}
\mathcal{A}(y)= \frac{1}{2} \epsilon(y) \delta(y)= \int_{0}^{y}  {dy' w_{0}(y' ) \delta '(y')/V },
\label{eq:melt}
\end{equation}

or in a differential form,  (after setting $\delta'(y)=\frac {2\epsilon(y)}  {\cos (\theta/2)} $ and $\delta(y)=2\epsilon(y) \tan (\theta/2) $)
\begin{equation}
w_{0}(y)= \frac{\partial \epsilon}{\partial y} V\sin{\theta}/2,
\label{eq:meltdiff}
\end{equation}
where $\epsilon(y)$ may be be written as the sum of two terms as $\epsilon(y) =\epsilon_{sk}(y)+\frac{\alpha(y)}{\sin(\theta/2)}$. In this formula $\epsilon_{sk}$ is the penetration depth of the blade at a given $y$ in the frame of reference of the skate and $\frac{\alpha(y)}{\sin(\theta/2)}$ is for the vertical height of the liquid layer, a layer which originates also from the melting of ice.  

\subsubsection{Length of contact and film thickness }

Using equations (\ref{eq:wo1})-(\ref{eq:poids})-(\ref{eq:meltdiff}) we can derive now the geometry and the components of the velocity in the film below the skate.  Let us first calculate the length of contact $\ell$ using equation (\ref{eq:poids}), that requires to express $\Delta w(y)$ and $\frac{\delta' (y)}{\alpha(y)}$. 

The  blades  of hockey and figure skates are curved along the propagation direction, in the plane $(z,x)$,  with a radius of profile $r$. 
 This radius of profile being much larger than the length  of contact (for hockey skating $r $ is  generally between $2m$ and $4m$),  the penetration depth of the blade writes 
 \begin{equation}
\epsilon_{sk}(y)= \frac{ y(2\ell-y)}{2r }.
 \label{eq:epsk}
\end{equation}

In a given cross-section the wetted length of the blade is $\delta'_{sk}=  \frac{ y(2\ell-y)}{r \cos(\theta /2)}$. Defining $\vert w_{sk}\vert =V \sin{(\theta/2)}\frac{\partial \epsilon_{sk}}{\partial y}$, which could be seen as  the  normal velocity of the blade with respect to an observer at rest (or the inverse), we have
 \begin{equation}
\vert w_{sk}(y) \vert = V\sin(\theta/2)\frac{ l-y}{r },
 \label{eq:wsk}
\end{equation}

and the flow velocity $w_{0} $ writes as 
 \begin{equation}
 w_{0}(y)= \vert w_{sk}(y) \vert +  V \frac{d\alpha}{dy},
  \label{eq:wo2}
\end{equation}

Introducing equations (\ref{eq:wsk})-(\ref{eq:wo2}) in Stefan relation (\ref{eq:wo1}), gives the differential equation for the film thickness $\alpha$ 
 \begin{equation}
\alpha \left( \sin(\theta/2 )\frac{\ell -y}{r } +\frac{d\alpha}{dy}  \right )= k ,
 \label{eq:wo3}
\end{equation}
which has to be solved with the initial condition $\alpha=0$ ahead of the blade (at $y=0$) because we have neglected the longitudinal squeezed flow. In (\ref{eq:wo3})  the length scale $k$ is given by the relation 
 \begin{equation}
k= \frac{\eta V}{2L \rho},
 \label{eq:k2}
\end{equation} 
a very small length in the case of lubricating melted water film and realistic velocities (equal to $1.8 10^{-11} m$ for  $V=12m/s$), therefore our study  takes place in the limits  
 \begin{equation}
k \ll  \alpha \ll \epsilon \ll \ell \ll  r .
 \label{eq:inegk2}
\end{equation}
As found below,  the film thickness $\alpha$  is of order $0.15\mu$, the penetration depth $\epsilon$ of order $10\mu$, the length of contact  $\ell$ of order $1 cm$, and the radius of profile  $r$ of order of a few meters.
Equation (\ref{eq:wo3})  is to be considered jointly with the set of equations (\ref{eq:wo1}) and  (\ref{eq:wsk})-(\ref{eq:wo2}), together with the condition  (\ref{eq:poids}) balancing the skater weight with the beneath pressure. This later set can be reduced to the integral relation, 

\begin{equation}
Mg=\frac{\eta V \sin(\theta/2)}{k^{3}} \int_{0}^{\ell}{  \left(\frac{2\ell y -y^{2}}{r \cos(\theta/2)} \right)^{3 }\left(\sin(\theta/2) \frac{\ell-y}{r } + \frac{d\alpha}{dy}   \right)^{4} dy}.
\label{eq:poids2}
\end{equation}
Our mathematical analysis finally yields the coupled set of equations (\ref{eq:wo3})-(\ref{eq:poids2}). This makes our main result and it will be studied  in the rest of the section.

 \subsection{ Solution of coupled equations  (\ref{eq:wo3}) and (\ref{eq:poids2}) }
 
 Equation (\ref{eq:poids2}) fixes the value of $\ell$ once the first equation  (\ref{eq:wo3}) is solved. In principle this makes a well defined problem. Since the equations are nonlinear, no analytical solution exists and a numerical solution must be looked at.  This can be done by iterations by taking successive values of $\ell$ until the procedure converges: choose  an arbitrary value  $\ell^{(n)}$ for the length of contact, solve the differential equation (\ref{eq:wo3}) which gives a solution $\alpha^{(n)}$,  insert  this solution in the r.h.s of equation (\ref{eq:poids2}) and increment $ \ell$  using the same procedure as before until condition (\ref{eq:poids2}) is satisfied.
However in the situation of ice skating, because the length scales are widely different, it makes sense to try to find the solution in the frame of the inequality (\ref{eq:inegk2}). 
 
  \subsubsection{ Solution at leading order }
  \label{sec:zero-order}

From  inequality  (\ref{eq:inegalfa}) one may conjecture that  equation  (\ref{eq:wo3}) has to be solved at leading order, by considering the derivative $\frac{d\alpha}{dy} $ as small in the term $\frac{w_{0}}{V}= \frac{\ell-y}{r }+\frac{d\alpha}{dy}$.  Defining

   \begin{equation}
 \kappa(y) = \frac{d\alpha}{dt}/ w_{0}=V\frac{d\alpha}{dy}/ w_{0},
    \label{eq:kappa}
\end{equation}

we get  the simple solution 
 
  \begin{equation}
\alpha^{(0)} (y) = \frac{ r  k}{ \sin(\theta/2)(\ell -y) )},
 \label{eq:wo3.1}
\end{equation}
in the limit $\kappa \ll 1$ ,
 where the superscript ${(0)}$ is to mean that this is the solution at leading order. Such an approximation requires to define a  small parameter  ($\gamma$, equation (\ref{eq:smallpara})) in terms of which the expansion of the solution is justified, at least in a certain domain. 
This solution has two obvious defects. It diverges at $\ell = y$ which contradicts (\ref{eq:inegalfa}) and makes infinite the derivative $\frac{d\alpha}{dy}$ which was assumed to be much smaller than $\frac{\ell -y}{r }$  to derive equation 
 (\ref{eq:wo3.1}) from equation (\ref{eq:wo3}).Therefore there is a boundary layer near $\ell=y$ (physically where the blade becomes exactly horizontal). The second defect of the leading order solution (\ref{eq:wo3.1}) concerns the  vicinity of $y=0$ where there is a second boundary layer, because at the front part of the ice-skate contact $\alpha(y)$  must tend to zero, a limit value obviously not satisfied by the solution  (\ref{eq:wo3.1}).

 However it turns out that, because 
 the two boundary layers have a small extension in the $y$ direction 
 one may neglect their contribution to the integral in equation (\ref{eq:poids2}), a result valid for all skating configuration as discussed in subsection \ref{sec:b.l.}. This integral which has to be calculated as well by canceling the derivative $\frac{d\alpha}{dy}$, becomes independent  of $\alpha(y)$ and can be computed. Using  the result $ \int_{0}^{\ell}  { dy(\ell-y)^{4} (2\ell y -y^{2})^{3}  }=\frac{16}{1155} \ell^{11}$,
we obtain 

\begin{equation}
 \ell^{(0)} =  \left( \frac{1155}{16 } C Mg \right)^{1/11}
\label{eq:ll3.1}
\end{equation}

with 
\begin{equation}
C=  \frac{ (\cos(\theta/2))^{3} } { \sin(\theta/2))^{5} }\frac {(\eta V)^{2}  r ^{7}}  { (2L\rho)^3}.
\label{eq:kl.1}
\end{equation}
This expression indicates that the length of contact is almost independent on the mass of the skater (as the power  $M^{1/11}$), depends on the velocity as $V^{2/11}$ and mainly depends on the radius of profile (as $r^{7/11}$).
Note that ultimately the  heat losses  have a quite small effect on the water layer geometry, because it just amounts to reduce the length of contact by a factor $2^{3/11} \sim 1.2$.

To close the leading order solution of the coupled set (\ref{eq:wo3}) and (\ref{eq:poids2}), one has to replace of $\ell$ by  $ \ell^{(0)}$ in (\ref{eq:wo3.1}). 
 Note that by plugging into this expression of $\ell ^{(0)}$ the numerical values for hochey skating, one finds that the inequalities (\ref{eq:inegk2})
 are satisfied.  In terms of the data, the approximation that the derivative $\frac{d\alpha}{dy}$ is negligible in equation (\ref{eq:wo3.1}), writes $\gamma \ll 1$ with 
   \begin{equation}
   \gamma = \frac{ k r ^{2} }{ \ell ^{3}},
   \label{eq:smallpara}
\end{equation}

because, by assuming  (\ref{eq:wo3.1}), we have $(\frac{ d\alpha}{dy})^{(0)} = -\frac{ r  k}{ \sin(\theta/2)(\ell -y) ^2)}$  which is of order $\frac{ r  k}{\ell^{2}}$ in the major part of the domain and the term $ \sin(\theta/2 )\frac{\ell -y}{r } $ is of order $\ell/r $ (the bite angle $\theta$ is a parameter which can be  taken away from the discussion of order of magnitude because  it  is about $\pi/2$).
 If the inequality $  k r ^{2} \ll \ell^{3}$ is satisfied,  the leading order solution (\ref{eq:wo3.1}) for $\alpha$  should be valid outside the boundary layers. This inequality is fulfilled for skating. This is not a consequence of $ k \ll  \ell \ll r   $, but follows from the order of magnitude typical of data, this ratio being of order  $10^{-4}$ for figure, hockey and speed skating. However the condition $ \gamma \ll 1$ does not guarantee that the leading order solution (\ref{eq:ll3.1}) is a good approximation for $\ell$, this point will be discussed in subsection \ref{sec:b.l.} where another condition, more drastic, is derived.

To go further of the understanding of the role the $y$-dependent ratio $\kappa(y)$
 we shall investigate the behavior of the solution close to the two boundary layers.

\subsubsection{Close to the boundary layers}
\label{sec:b.l.}
 Close to $y=\ell$ and $y=0$  local forms of equation  (\ref{eq:wo3}) can be analytically derived by using  standard asymptotic expansion of a solution of a differential equation with a small parameter. This requires to define specific scalings valid in regions much smaller than $\ell$.   
 
 1) At the front of the blade, in the limit $y \to 0$,  the solution  of equation  (\ref{eq:wo3}) with i.c. $\alpha(0)=0$ is $ \alpha(y) = (2k y)^{1/2},$  which  has a square root singularity.  In a certain domain  one can  neglect  $y$ with respect to $\ell$ in (\ref{eq:wo3}). This leads to define, in this domain, the scaled variables  $\tilde{y} =\frac{ y}{\ell_{0} }$ and  $\tilde{\alpha} = \frac{\alpha}{\alpha_{0}}$  with  the length scales $\ell_{0} =\frac{k r ^2}{\ell^2} $   and $\alpha_{0}=\frac{k r }{\ell} $, a scaling  consistent  with the use of the lubrication approximation for the liquid layer because $ \ell \ll r  $. In the intermediate domain which makes the transition between the solution $ \alpha(y) = (2k y)^{1/2}$ and the exact one, equation (\ref{eq:wo3}) takes the local form
\begin{equation}
\tilde{ \alpha}  \left( \sin(\theta/2 ) +\frac{d \tilde{\alpha}}{d \tilde{y}}  \right )= 1.
\label{eq:cl2}
\end{equation}

Its solution 
behaves asymptotically like 
$$\tilde{ \alpha}  \approx \frac{1}{\sin(\theta/2 )} $$ which matches the leading order solution  (\ref{eq:wo3.1}) far away from the front part of the skate in these scaled variables.  Restoring the original variables, the solution $\alpha^{b.l.}(y)$ of the local equation (\ref{eq:cl2}) is undistinguishable from the exact one for $0 \le y \le \ell_{0} $, as illustrated by the insert of Fig. \ref{fig:front}. Further the local solution tends to the constant  value $ \alpha_{0} =  \frac{k r }{\ell} $. The range  $\ell_{0} << y << \ell$   defines the matching domain.  At larger values of $y$, the local solution remains constant whereas the exact solution $ \alpha(y)$ remains close to $ \alpha^{(0)} (y) $ until the second boundary layer.

2) At the end of the contact, where $y$ tends to $\ell$, the leading order approximation is not valid, as written above, because the solution (\ref{eq:wo3.1})  is derived under the condition that the derivative $\frac{d\alpha}{dy} $ is small, whereas this derivative  diverges at $ y =  \ell$ (as well as $\alpha$).
Therefore we have to consider that all terms of equation (\ref{eq:wo3}) become of the same order of magnitude, that 
 leads to introduce the following length scales

  \begin{equation}
  \alpha_{*}=  (k^{2}  r )^{1/3};    \; \; \; \,\; \; \; \     \ell_{*}= (k  r ^{2})^{1/3}; 
\label{eq:bl3}
\end{equation}

which gives in particular the order of magnitude of the thickness in this domain.  Notice that because  $ k\ll r  $ the length scale $\alpha_{*}$ is much smaller than  $\ell_{*}$ as necessary for maintaining the validity of the lubrication approximation for the flow in the layer between the skate and ice.  

By introducing $\overline{y} =\frac{ \ell-y}{\ell_{*} }$ and  $\overline{\alpha} = \frac{\alpha}{\alpha_{*}}$ equation  (\ref{eq:wo3}) is transformed into 
 \begin{equation}
\overline{\alpha}  \left( \sin(\theta/2 )\overline{y}- \frac{d\overline{\alpha} }{d\overline{y}}  \right )= 1 ,
 \label{eq:wo3.2}
\end{equation}

which is  in fact identical to the original one, up to the  rescaling defined in (\ref{eq:bl3}), and the change of variable from $\ell -y$  
to  $y$. Therefore in this case the local solution is the same as the global one and we cannot derive any analytical solution or numerical one
from initial conditions taken at $\overline{y}=0$  ($y=\ell$), because the forward solution $\overline{\alpha} $ of the differential equation (\ref{eq:wo3.2}) is unstable  (as well as the backward  solution  $\alpha(y)$ of (\ref{eq:wo3})). This result could appear without any interest, but it is not, because the relation (\ref{eq:bl3}) is quite relevant, since it gives the order of magnitude of all the unknown lengths without  solving any differential equation, in agreement with the exact numerical results, as reported next.

 \subsection{Numerical results}
\label{sec:numerics}
In the numerical study we take the following data  (in SI units)  $\eta = 10^{-3}\;  kg/(m . s)$, $L =  3.3 \; 10^{5} \; J/Kg$, $g= 10 \;m/s^2$ and  for the skater $V = 12 \;m/s$, $M = 75 \; kg$ (except  in the subsection devoted to the  role of these parameters). Moreover  we set  respectively  {$r =3m$  and $r =25m$  depending on wether we  describe  figure/hockey  skating or speed skating (the radius of profile being between about $2m$ for figure skating, between $2 m$ and $4 m$ for hockey, and may be longer than $25m$ for speed skating).
In all cases we show  that
the  assumption $\gamma \ll 1$, made to guarantee that the parameter $\kappa(y)$ is small in the major part of contact, is fulfilled. More importantly we are going to show that the  leading order solution $\ell^{(0)}$ is a good approximation for $\ell$, that  allows to shorten the numerical calculus of the solution of the two coupled equations by making a single iteration. 

\subsubsection{Figure and Hockey skating}

The exact solution of the coupled equations requires to increment $\ell$ step by step, using the method outlined above: solve  (\ref{eq:wo3}), calculate the r.h.s. of  (\ref{eq:poids2}) until it is equal to its l.h.s. We find  that the resulting length of contact is very close to $\ell^{(0)}$. Therefore a short cut  to get a solution in good agreement with the exact one is to  use this leading order value  for $\ell$ and solve (\ref{eq:wo3}) only once. Doing this 
 the solution of equation (\ref{eq:wo3}) with initial condition $\alpha(0)=0$  is found to agree very well with the exact one (the error being less than one per thousand for the film thickness), 
  so that they are indistinguishable in fig. \ref{fig:final}, both merging in the red solid line. The numerical results obtained with the single step integration of (\ref{eq:wo3})) are
 
  \begin{equation}
  \alpha^{(1)}(\ell)= 0.159 \mu ;    \; \; \; \, \epsilon^{(1)}(\ell)= 20.95 \mu    \; \; \; \,    \; \; \; \, \ell^{(0)}= 1.12 cm; 
\label{eq:num2}
\end{equation}
 in very good agreement with the exact values, 
 \begin{equation}
  \alpha(\ell)= 0.159 \mu ;  \; \; \; \  \epsilon(\ell) = 20.99 \mu;  \; \; \; \,    \; \; \; \,  \ell= 1.12 cm,
\label{eq:num1}
\end{equation}

obtained by the multiple step method outlined above. These lengths are ordered as required by equation (\ref{eq:inegk2}),  moreover we have $w_{0} \ll V$, as expected. 

   \begin{figure}
{
(a)\includegraphics[height=1.5 in]{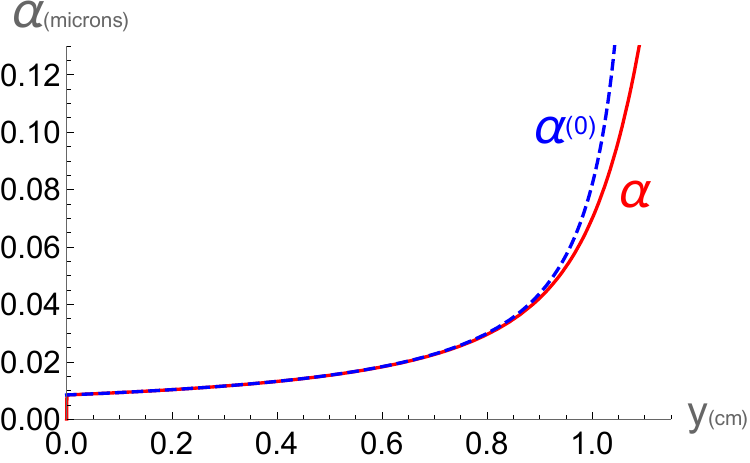}
(b)\includegraphics[height=1.5 in]{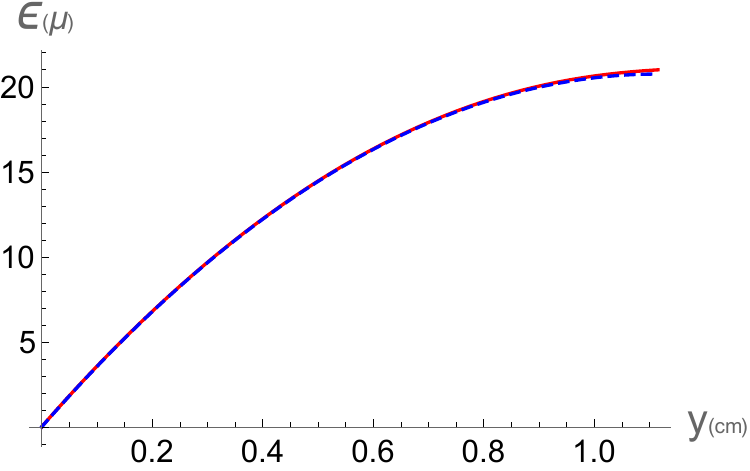}

}
\caption{ (a) Film thickness $\alpha(y)$ and (b) penetration depth $\epsilon(y)$. The leading order solution (blue-dashed line) is  compared to the  exact solution of equations (\ref{eq:wo3}) -(\ref{eq:poids2}) with i.c. $\alpha(0)=0$, drawn in red-solid line which is indistinguishable from the first order one.}
\label{fig:final}
\end{figure}

For comparison the leading order solution is shown in blue dashed line. In (a) $\alpha^{(0)}(y)$  agrees  with the exact one over the main part of the bottom blade, but differs in front close to $y=0$  (hardly visible in this figure but detailed in fig.\ref{fig:front}) and close to $y=\ell$, as stated previously (subsection \ref{sec:zero-order}).
To illustrate the validity of the conjecture  made (at the beginning of the previous subsection) that  the derivative $\frac{d\alpha}{dy} $ is small compared to $\frac{w_{0}}{V }$, we compare these two quantities in Fig. \ref{fig:wo-ap}.  In this figure the derivative term is enhanced by a factor ten to be visible. The dashed curve, drawn for  $10 \frac{d\alpha}{dy} $, shows that the derivative term is at least one order of magnitude lower than $\frac{w_{0}}{V }$ (solid line), except in the vicinity of  $y=0$ and $y=\ell$, as expected. This result shows the importance of taking account of the relative velocity of the skater with respect to the ice surface, when writing the expression of the melting rate, as yet noted.

   \begin{figure}
\centerline{ \includegraphics[height=1.5 in]{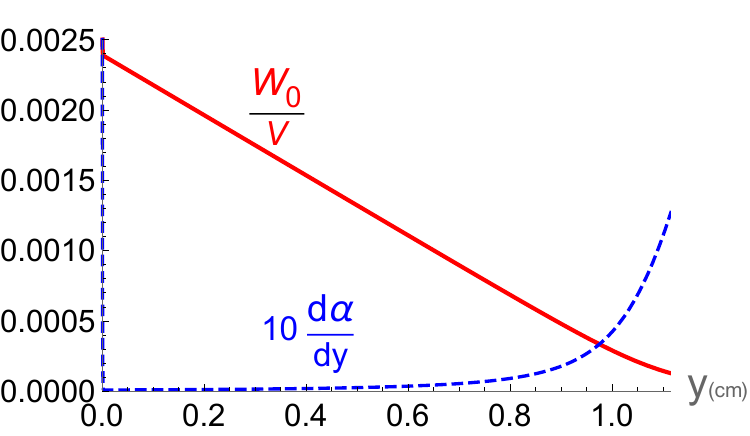}}
\caption{  $\frac{ w_{0} }{V }$ (red solid line) and $\frac{d\alpha}{dy} $ versus $y$ (blue dashed line)}.
\label{fig:wo-ap}
\end{figure}

In the two boundary layers the numerics agree with the analytical study presented just before. 
Close to the origin  Fig. \ref{fig:front} shows the  solution of the local equation (\ref{eq:cl2})  together with the exact one. They merge including in the matching domain which extends on a distance few times $ \ell_{0}$ where $\ell_{0}=\frac{k r ^2}{\ell^2}= 1.5 \mu$. 
The insert displays the whole domain $0 \le y \le \ell$  to show that the local solution is no more valid beyond the matching domain where it remains constant.
  
The second boundary layer, close to $y=\ell$ , is clearly visible in Fig.\ref{fig:final}-(a) because  the leading order solution $\alpha^{(0)}$ is well separated from  the exact one  over a  noticeable distance. The extension of this layer, of order $\ell_{*}=0.05 cm$ (the difference between the abscissa of the two curves in the right of the figure), is much larger than the extension of the boundary layer in front. We point out that the analytical study of the boundary layer presented above allows to derive straightforwardly 
 the order of magnitude of the thickness in terms of the data. Equation (\ref{eq:bl3}) gives $ \alpha_{*}=0.11 \mu$  in  qualitative agreement with 
 the exact result,  a result obtained without solving the  coupled set of equations (\ref{eq:wo3})-(\ref{eq:poids2}). 

   \begin{figure}
\centerline{ \includegraphics[height=2in]{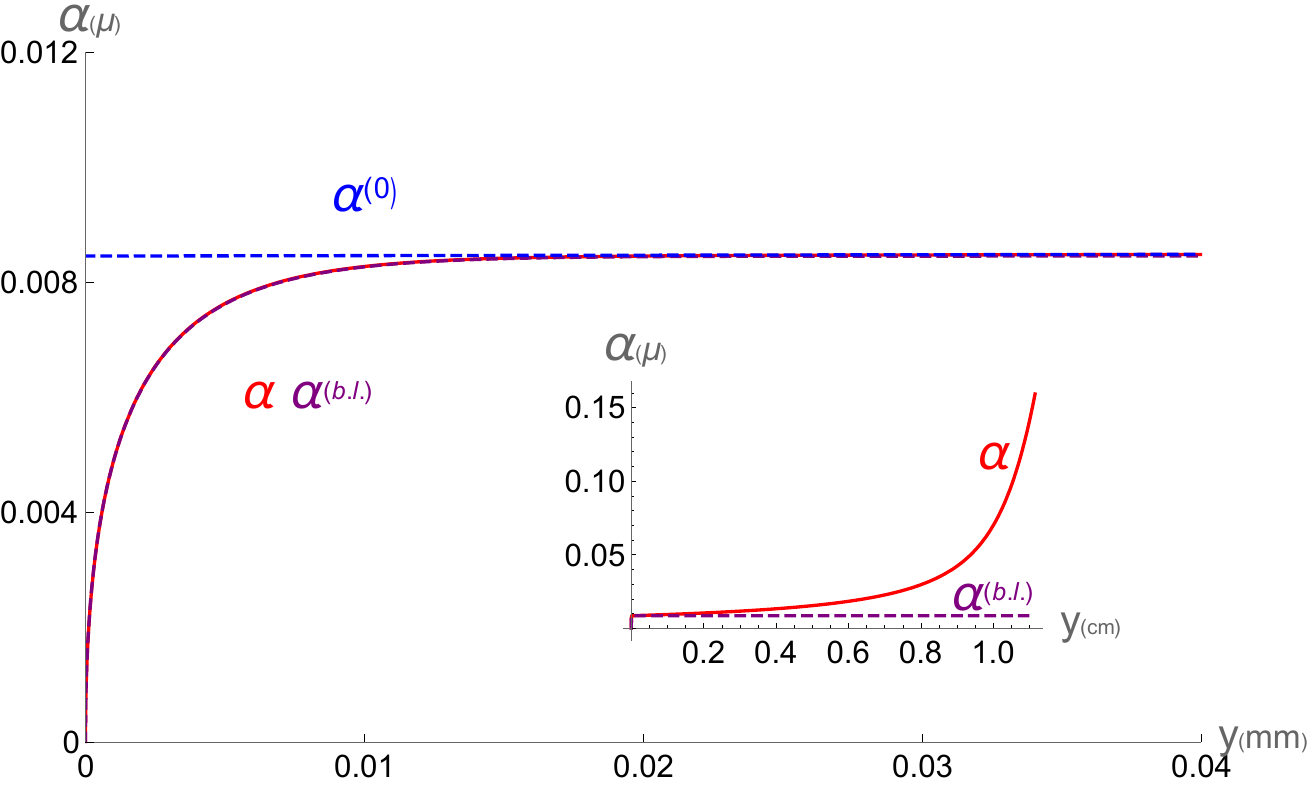} }
\caption{        Boundary layer description close to $y=0$: the exact solution (red solid line) is compared  with the local solution of  equation (\ref{eq:cl2}) $\alpha^{b.l.}$ (in restored variables, plotted in purple dashed line). They merge in the principal curve drawn  in the domain $0 \le y \le 27 \ell_{0}$ where $\ell_{0}=\frac{k r ^2}{\ell^2}= 1.5 \mu$.
 The insert shows the solutions over the full domain to show that the matching domain (defined in the text) is much shorter than $\ell$.    }
\label{fig:front}
\end{figure}

To compare our theory with the one presented in references  \cite{LScana}-\cite{LScana2}, we note that they differ on several points. First the condition (\ref{eq:meltdiff}) necessary to describe the formation of the trough is lacking in this study, that leads them to estimate the length of contact from its value at rest, whereas the dynamical length of contact obtained here is five times  larger. Secondly the authors of  \cite{LScana}-\cite{LScana2} identify the melting rate $w_{0}$ with the thickness growth rate $\frac{d\alpha}{dt}$ although  we  show that it is more than one order of magnitude larger almost everywhere below the blade. This result shows the importance of taking into account the relative motion of the ice with respect the skater when writing the expression of the melting rate $w_{0}$ in the frame of the skater. Lastly the authors of  \cite{LScana}-\cite{LScana2} add/subtract linearly the effects of the different contributions to the thickness growth rate, that leads to a  \emph{single} differential equation, although  our treatment  leads to a set of  \emph{two} coupled equations, one for the film thickness \textit{and} the  other for length of contact. 
 
  \subsubsection{Role of the parameters}
  \label{sec:data role}
  
It could be interesting to know the optimal conditions of skating. To give an idea of the role of the various parameters on the film thickness, let us investigate the role of the parameters which could be changed,  like  the mass $M$ and  velocity $V$ of the skater,  and the sharpening of the blade (its radius of profile $r $ and bite angle $\theta$). Here we investigate the role of  these external parameters,  keeping unchanged the other ones.  The result is summarized  in fig. \ref{fig:roleUr}. 

   \begin{figure}
\centerline{   
 (a)\includegraphics[height=1.2 in]{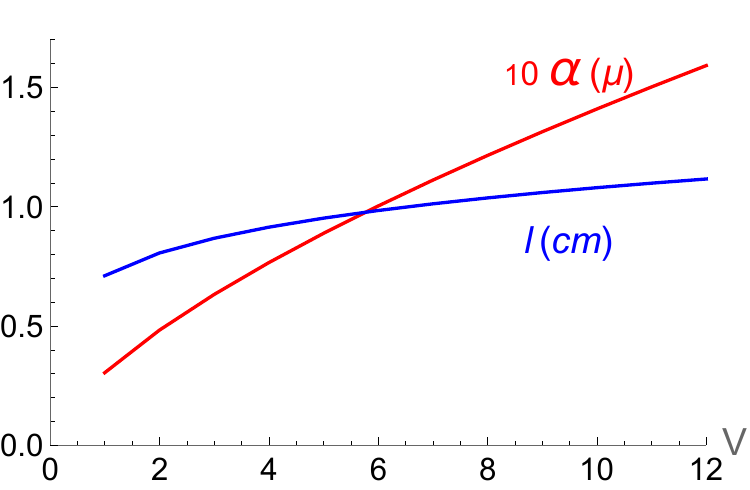}
(b)\includegraphics[height=1.2 in]{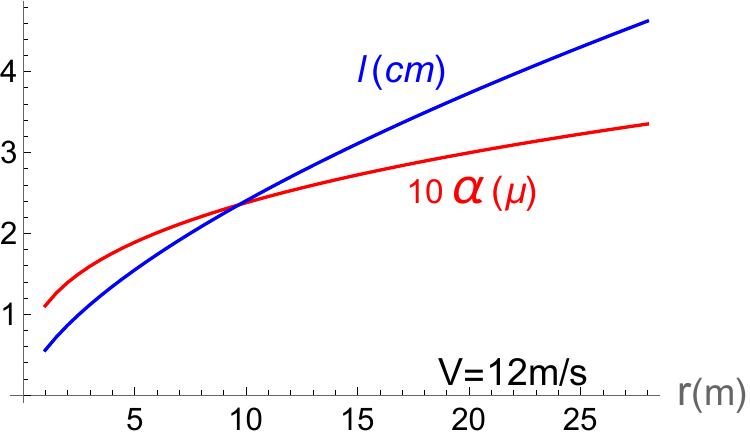}  }
\centerline{
(c) \includegraphics[height=1.2 in]{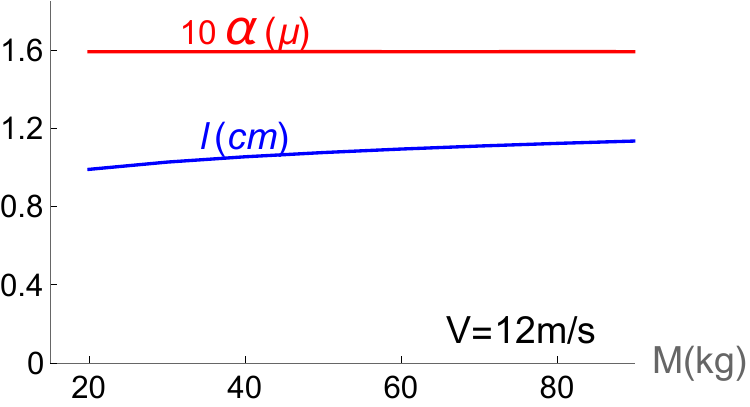}  
(d) \includegraphics[height=1.2 in]{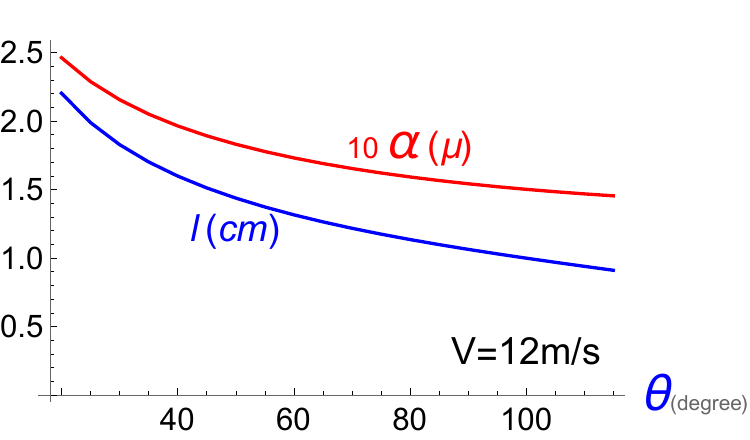}  }

\caption{ Film thickness $\alpha(\ell)$ (multiplied by $10$) in microns, and length of contact $\ell$  in cm, versus the following parameters:  (a)  skater velocity $V$ (all other parameters are fixed), (b)  radius of profile of the skate, (c) skater mass  and (d)  bite angle of the blade; the  fixed parameters are identical to those of previous figures.  }
\label{fig:roleUr}
\end{figure}

 Our curves show clearly that the film thickness and also the length of contact strongly depend  on the velocity of the skater,  moderately depends on the radius of profile and on the bite angle, although the geometry of the layer is almost insensitive to the skater mass.  
Let us now compare these numerical results with the analytical study of solution of sec. \ref{sec:b.l.}. Introducing  the bite angle dependence, the order of magnitude of the film thickness  (at $y \backsimeq \ell$) given by (\ref{eq:bl3}) becomes,
 \begin{equation}
 \alpha_{*}= (\frac{ \eta }{ 2L })^{2/3}  \left(  \frac{ r V^{2}  }{ \sin{(\theta /2)} } \right)^{1/3}.
 \label{eq:num-para}
 \end{equation} 
  This relation is in qualitative agreement with all numerical results displayed in fig.\ref{fig:roleUr}. It predicts a stronger dependence versus $V$ than  versus $r $ and $\theta$, and independence with respect to  $M$. Therefore  we can claim that it contains  the main physical result of the present study.

 \subsection{Speed skating with an inclined blade}
\label {sec:speed}

The model derived above can be applied to  the case of a speed skater  strongly inclined with respect to the vertical axis, this angle being equal to $\pi/4$.  A more general formulation can be derived in the general case of any angle value, but not presented here. For an inclined and symmetric rectangular blade , the geometry of the contact region is very close to the one considered  for hockey or figure skating,  but has to be calculated for larger  radius of profile  value and setting $\theta=\pi/2$.  The radius of profile for speed skating is almost ten times larger than for figure/hocker skating, its typical value being  $25m$,  although the value of $\theta$ is close to the typical values of hockey-skating  for which bite angles are between $70$ and $85$ deg.

 Because  the derivation of Stefan flow is the same for an inclined blade as for a V-shaped blade, it is enough to insert these two different parameters $ \theta, r $ into the equations obtained in section \ref{sec:hockey}. We get the following results ,
 \begin{equation}
  \alpha(\ell)= 0.4 \mu ; \,\,\,  \epsilon(\ell) = 66 \mu; \, \, \,    \ell= 5.73 cm.
\label{eq:numspeed-V}
\end{equation}
This shows the role of the radius of profile: increasing it by a factor $8$ increases the depth of the through and the thickness of the film by a factor $3$.  Except this factor, the geometry  of the trough, namely  the functions $\epsilon[y]$ and $\alpha[y]$ evolves  similarly as  Fig. \ref{fig:final}  where the exact solution of the coupled equations  (\ref{eq:meltdiff})-(\ref{eq:poids2})  are very close to the first order solution obtained by using (\ref{eq:ll3.1})   for solving  (\ref{eq:wo3}).  In summary we show that speed-skaters can plough a trough three times larger and deeper than hockey or figure skaters when skating with an inclined blade at the same velocity, and that the film of water is also three times thicker. These results are completely different from the study of \cite{LScana2} where the film thickness is of order $1 \mu$, whatever the angle of the blade.


\subsection{ Critical mass}
\label{sec:adim}

The analytical study of the coupled equations   (\ref{eq:wo3}) -(\ref{eq:poids2})  has enlightened  the role  of the quantity  $ \kappa(y)$  defined in (\ref{eq:kappa}). We have shown that this quantity  can be neglected to get  the solution of the integral equation (\ref{eq:poids2}): this gives $\ell^{(0)}$ in very good agreement with the exact length of contact, although $\kappa$ cannot be neglected to get the solution of the differential equation (\ref{eq:wo3}), because it  becomes of order unity, as discussed in section \ref{sec:b.l.} and illustrated in Figs.\ref{fig:final}(a)-\ref{fig:wo-ap}-\ref{fig:front}. 
There remains a question: What is the domain of validity of the approximation $\ell \backsimeq \ell^{0}$ ? In other words is there a range of parameters such that the leading order solution $\ell^{(0)}$ differs noticeably from the exact result? To answer this question we shall quantify the contribution of $\kappa$ in the integral equation (\ref{eq:poids2}) by using the results of subsection \ref{sec:b.l.}. Let us consider the integral in the r.h.s. of (\ref{eq:poids2}) 

\begin{equation}
\mathcal{I}=\int_{0}^{\ell}{ (\delta'_{sk})^{3} \left(\frac{w_{0}(y)}{V} \right) ^4}dy,
\label{eq:b1}
\end{equation} 
and set $\Delta \mathcal{I}= \mathcal{I}-\mathcal{I}^{(0)} $ where $\mathcal{I}^{(0)}$ is for  $w_{0}= w_{0}^{(0)}$. The condition $\frac{\Delta \mathcal{I} }{ \mathcal{I}} \ll 1$  is the one we are looking for, because it means that $\ell^{(0)}$ is a good approximation of $\ell$. The integrand in $\Delta \mathcal{I}$ is significant only in the region $y\sim \ell$ of extent $\ell_{*}$. Assuming that $\kappa \sim 1$ in this domain, and ignoring the role of the bite angle $\theta$ we obtain 
$$\Delta \mathcal{I}=15(\frac{\ell_{*}}{r })^{4}\ell_{*}(\frac{\ell^{2}}{r })^{3}, $$  moreover we have $ \mathcal{I} \sim \frac{16}{1155}\frac{\ell^{11}} {r^{7} }$ at leading order. In terms of the length scales,
the condition  $\frac{\Delta \mathcal{I} }{ \mathcal{I}} \ll 1$ can be written as
 \begin{equation}\,
 \frac{\ell_{*}}{\ell }  \ll 0.25 ,
 \label{a40}
 \end{equation}
 a condition fulfilled above (the ratio $\frac{\ell_{*}}{\ell }$  is equal to $10^{-4}$ in the case  of Fig.\ref{fig:final}). To compare with the inequality (\ref{eq:smallpara}) which is for the condition  $\kappa \ll1$ in the central domain, we may notice that $\gamma= ( \frac{\ell_{*}}{\ell } )^{3}$, therefore the condition (\ref{a40}) is more drastic than (\ref{eq:smallpara}). This is because for deriving (\ref{a40}) we  make one step more since we  impose that the contribution of $\kappa$ in the boundary layer is negligible.
In terms of the data, using the order of magnitude (\ref{eq:bl3})  for $\ell_{*}$ and equation (\ref{eq:ll3.1}) for $\ell$, the condition (\ref{a40})  allows to define a set of critical parameters for the mass or velocity of the skater, and for the radius of profile of the blade. For example the inequality (\ref{a40})  writes $M \gg  M_{*}(V,r)$ with

  \begin{equation}
M_{*} \sim  c_{m} \frac{2L \rho }{g}\left( r k^{5}  \right)^{1/3}
\label{eq:a4}
\end{equation}
where $M_{*}$  has the dimension of a mass, $c_{m}$ is a numerical coefficient ,  $c_{m}= (16/1155) (\cos\theta)^{5}/ [(\sin \theta)^{3}  0.25^{11}]$  and the length $k$ is proportional to the speed $V$.  Using the data taken above for hockey skating with $V=12m/s$, $r=3m$, this critical mass  $M_{*}$ is  very small ($2 \mu g$).  To get critical mass values of order one kg, the velocity should be increased by a factor $2500$ which would be inaccessible for skaters. This result shows that 
  the relation (\ref{eq:bl3}) covers all realistic skating situations.

From the previous estimates, one can also derive the friction force on the skate of length $l$. This is the friction on a surface $ l  \delta'$ in a fluid where the velocity gradient is of order $V/ \alpha$. This total force (not per unit length of the skate) is $$ F_v =  \frac{\eta V l \delta'}{\alpha} \mathrm{,}$$ where  $\delta' $ and $\alpha$ have been estimated above.  A dimensionless measure of this friction is the ratio of $ F_v$ to the weight $ M g $ of the skater 
With the data taken above, this ratio is about  $10^{-4}$.

As a final remark, let us note that the geometry considered in this section, namely a narrow furrow made by melting ice, solves the problem of turning, because by tilting at the right angle, the skater rests on an inclined furrow which can stand a priori both his/her weight and the centrifugal force generated by the turn.

\section{Rectangular blade: Vertical speed skating}
\label{sec:rectangular}
We keep the same notations as before and assume that the skater moves on a vertical (not inclined) blade, with a rectangular cross-section of width $\delta$  independent of $y$. The blade is sunk into a furrow of  total depth $\epsilon (y)= \epsilon_{sk}(y) + \alpha(y)$, where $\alpha$ is the thickness of the  film of melted water. To simplify the calculations we assume that the thickness is constant all along the cross section, namely that $\alpha$ only depends on $y$ as above, whereas the value of $\alpha(s,y)$ could be larger vertically than  horizontally where the film is squeezed by the weight of the skater. Then the furrow has a total depth $\epsilon (y)$, total width $\delta + 2 \alpha (y)$, and the length  of the ice boundary at the abscissa $y$ is $ \delta'(y)= \delta +2 \epsilon(y)$. 

Equation (\ref{eq:wo1}) which describes the  balance between the heat  dissipated by viscous friction and absorbed to melt the ice surface  is  still valid, so are Poiseuille equations (\ref{eq:dp1})-(\ref{eq:u1}) and continuity condition  (\ref{eq:u2}), together with equations  (\ref{eq:d2p})-(\ref{eq:deltaw}). The relation between the melting rate $w_{0}$ and the slope of the skate blade is deduced as above by  writing the condition for the formation of the trough,
$\mathcal{A}(y)= \epsilon(y) \delta= \int_{0}^{y}  {dy' w_{0}(y' ) \delta '(y')/V }$, that gives after  derivation with respect to $y$,

\begin{equation}
w_{0}(y)=  V\frac{\delta}{\delta +2 \epsilon(y)} \frac{\partial \epsilon}{\partial y}.
\label{eq:melt2}
\end{equation}

The drop of pressure (\ref{eq:pr}) is also the solution   of (\ref{eq:d2p}) (with the boundary conditions as in previous section),  The integration of the pressure over the surface of contact,  balanced with the weight of the skater writes here

\begin{equation}
Mg=\eta\delta \int_{0}^{\ell}{  w_{0}(y) \frac{ K_{1}(y)}{\alpha^{3}(y)} dy},
\label{eq:poids3}
\end{equation}

where $K_{1}(y)= \delta^{2} +6 \left(\epsilon^{2}(y) +\epsilon(y)\delta \right) $.

Finally, setting  $\epsilon=\epsilon_{sk} +\alpha$ in (\ref{eq:poids3})  and (\ref{eq:melt2}),  the two coupled equations relating the thickness $\alpha(y)$ to the length of contact $\ell$ are

\begin{equation}
\alpha \left( \frac{\ell -y}{r } +\frac{d\alpha}{dy } -2\frac{k}{\delta} \right )= k (1+ \frac{2\ell y-y^{2}}{\delta r })
\label{eq:wo4}
\end{equation}

which replaces (\ref{eq:wo3})
 and
 
 \begin{equation}
M g = \frac{V \eta\delta}{k^{3}}  \int_{0}^{\ell}  { dy \left(\frac{w_{0}}{V}\right)^{4}  K_{1}(y)},
\label{eq:poids4}
\end{equation}

which replaces (\ref{eq:poids2}), with 
 \begin{equation}
\frac{w_{0}}{V}= \frac{\delta(\ell-y)+ \frac{d\alpha}{dy}r }{\delta r + (2\ell y-y^{2}) +2r \alpha(y) } 
\label{eq:poids4b}
\end{equation}

and $$K_{1}(y)= \delta^{2} +6\left( (\frac{y\ell-y^{2}}{\ell})^{2} + \delta \frac{y\ell-y^{2}}{\ell} \right)$$

\subsubsection{Length of contact at leading order}
Neglecting the film thickness and its derivative in the above equations (\ref{eq:wo4})-(\ref{eq:poids4b}) gives the leading order solution $\frac{w_{0}^{(0)}}{V}=  \frac{\delta}{\delta + \epsilon^{(0)}}\frac{d\epsilon^{(0)}}{ dy} $. With 
\begin{equation}
\epsilon^{(0)}=(2\ell^{(0)} y -y^{2})/r ,
\label{eq:epspeed}
\end{equation}
  we obtain $\frac{w_{0}^{(0)} }{V}=\frac{\ell-y}{r}\frac{1}{1+\frac{2\ell y-y^{2}} { r\delta }}$ and
\begin{equation}
\alpha^{(0)} (y)=k  \frac{  \delta r  +  (2\ell^{(0)} y-y^{2})  }{  (\ell^{(0)} -y)\delta   }
\label{eq:alfospeed}
\end{equation}

and a length of contact solution of the integral relation

\begin{equation}
\int_{0}^{\ell^{(0)}}  { dy  (\frac{w_{0}^{(0)}}{V})^{4} \left( \delta^{2} +6(\frac{ 2\ell^{(0)} y-y^{2} }{ r  })^{2} +6\delta \frac{2\ell^{(0)}y-y^{2}}{r }  \right)   } = Mg\frac{k^{3}} {V \eta\delta} 
\label{eq:mgospeed}
\end{equation}
which can be solved by iteration.
The numerical result is $\ell^{(0)}=0.38cm$, and   curves $\alpha^{(0)} (y)$ and $\epsilon^{(0)}(y)$, which diverge at $y=\ell$ as in the case of V-shaped blade studied above. These curves are the blue-dashed ones  shown in Figs. \ref{fig:speed-flat}.
   \begin{figure}
{
(a)\includegraphics[height=1.5 in]{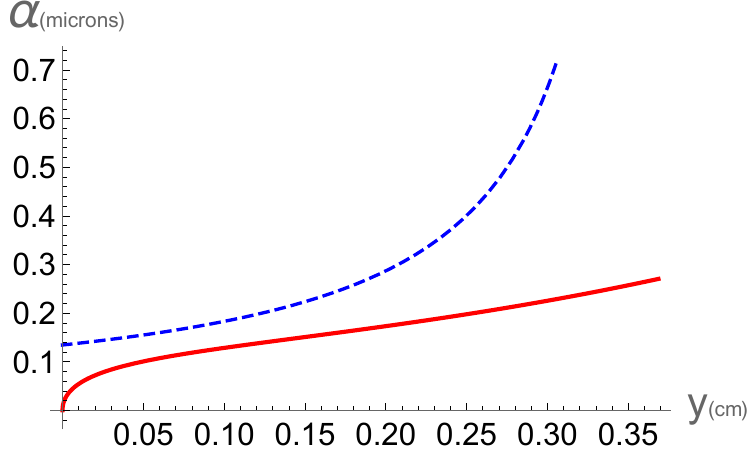}
(b)\includegraphics[height=1.5in]{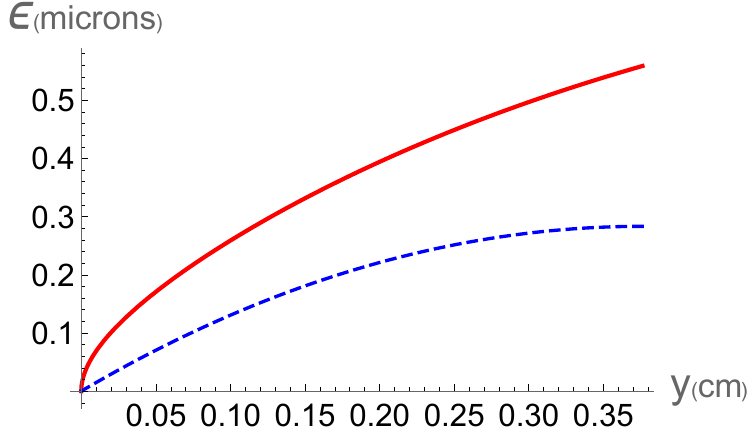}
}
\caption{ Film thickness and penetration depth in the case of speed skate  straight on ice (rectangular bottom blade) (a) Film thickness $\alpha(y)$ , (b) penetration depth $\epsilon(y)$. The dashed lines display the leading order solution, the solid red  curves is the first order one.
The leading order solution  obtained by solving equations ((\ref{eq:epspeed})-(\ref{eq:alfospeed})-(\ref{eq:mgospeed}), the first order solution is get by insering the leading order solution into (\ref{eq:melt2})-(\ref{eq:poids3}) .}
\label{fig:speed-flat}
\end{figure}

\subsubsection{First order solution}
Inserting the leading order solution into the exact equations (\ref{eq:melt2})-(\ref{eq:wo4}), we get first order solution. The film thickness $\alpha^{(1)}(y)$  and the depth of the through $\epsilon^{(1)}(y)$ are plotted in red solid lines in Figs.\ref{fig:speed-flat}.  We observe that the convergence of the solution is far from being obtained at first order. Moreover there is no possibility to pursue the iteration process  at second order because  at next order equation (\ref{eq:poids3}) has no solution. The explanation lies in the fact that  depth of the through $\epsilon^{(1)}(y)$   is only twice the film thickness, we get

 \begin{equation}
  \alpha(\ell)^{(1)}= 0.27 \mu ; \,\,\,  \epsilon(\ell) ^{(1)}= 0.56\mu; \, \,  \; \, \,    \ell^{(0)}= 0.38 cm,
\label{eq:numspeed}
\end{equation}

that doesn't  agree with the basic hypothesis  of our model which are summarized in (\ref{eq:inegk2}). Therefore another model has to be found (not done here).

\section{Summary and conclusion}
The problem of understanding how and why it is possible to skate on ice almost without friction has remained imperfectly understood for a long time.  In this work we introduced and discussed in depth all the necessary ingredient of a coherent theory. In particular we emphasized that, for a curved wedge-like blade, the one used in all skating sports, either speed, hockey or figure skating, the geometrical parameters and the velocity of the flow are linked together by three rather complex relations. One  describes the melting of ice by viscous friction in the thin layer of melted ice (Stefan condition), another one expresses that direct contact between the skate and ice is avoided because the pressure of the Poiseuille flow in the layer is sufficient to lift the weight of the skater above the ice surface, the third one is the kinetic condition which connect the volume of melted ice to the volume of the furrow dug by the blade. Putting together those three relations we have shown that all the unknown physical quantities can be deduced from a set of two coupled equations, one differential the other integral, for the film thickness $\alpha$ and the length  of contact $\ell$. The analytical study of both equations has revealed the existence of a  parameter ($ \kappa = \frac{\frac{ d\alpha }{dt}}{ w_{0}} $) which plays a key role in the derivation of the solution because it is small all along the film, except at the ends. This property shows in particular that the melting rate $w_{0} $  definitely hasn't to be confused with the  growth rate of the layer $\frac{d\alpha}{dt}$, an assumption found in the literature, because the two quantities differ by one order of magnitude in the correct theory. Besides this remark  the fact that  such a small parameter  exists is of prime importance from a practical point of view, because it allows to get analytically  the order of magnitude of the solution  and inform about the effects of the input data.

A natural extension of this complete theory is to allow the design of improved skates by optimizing their shape to lower friction and increase the grip on the ice surface when the skater makes a turn. 
\section{Acknowledgment}
We greatly acknowledge Christophe Clanet and Caroline Cohen for their interest in this work, and  stimulating discussions.
One of us (YP) would like also to thank his grand son, Gaspard, for teaching him patiently the ways of hockey skating.

      \end{document}